\documentclass[runningheads]{llncs}
\usepackage[T1]{fontenc}
\usepackage{graphicx}
\begin{document}
\title{Branch Learning in MRI: More Data, More Models, More Training}
%
%
\author{Yuyang Li\inst{1}\orcidID{0009-0005-5532-199X} \and
Yipin Deng\inst{1} \and
Zijian Zhou\inst{1} \and
Peng Hu\inst{1}\thanks{Corresponding author.}}
\authorrunning{Y.~Li et al.}
\institute{ShanghaiTech University, 393 Huaxia Middle Road, 201210 Shanghai, China\\
\email{\{liyuyang,dengyp2024\}@shanghaitech.edu.cn}\\
\email{hupeng@shanghaitech.edu.cn} \\
\url{https://github.com/5o1/Moero}
}
\maketitle              
\begin{abstract}
We investigated two complementary strategies for multicontrast cardiac MR reconstruction: physics-consistent data-space augmentation (DualSpaceCMR) and parameter-efficient capacity scaling via VQPrompt and Moero. DualSpaceCMR couples image-level transforms with kspace noise and motion simulations while preserving forwardmodel consistency. VQPrompt adds a lightweight bottleneck prompt; Moero embeds a sparse mixture of experts within a deep unrolled network with histogram‑based routing.

In the multivendor, multisite CMRxRecon25 benchmark, we evaluate fewshot and out-of-distribution generalization. On small datasets, k‑space motion‑plus‑noise improves reconstruction; on the large benchmark it degrades performance, revealing sensitivity to augmentation ratio and schedule. VQPrompt produces modest and consistent gains with negligible memory overhead. Moero continues to improve after early plateaus and maintains baseline‑like fewshot and out‑of‑distribution behavior despite mild overfitting, but sparse routing lowers PyTorch throughput and makes wall clock time the main bottleneck. These results motivate scale‑aware augmentation and suggest prompt‑based capacity scaling as a practical path, while efficiency improvements are crucial for sparse expert models.

\end{abstract}
\section{Introduction}

Cardiac magnetic resonance (CMR) imaging is a vital non-invasive tool for assessing cardiac structure and function, offering exceptional soft tissue contrast and radiation-free imaging. Multi-contrast CMR sequences, such as Cine for heart motion, Aorta for vascular assessment, Mapping for tissue characterization, and Tagging for strain analysis, are essential for accurate cardiovascular diagnosis. However, its extended scan times, often over 30 minutes, increase patient discomfort, risk of motion artifacts, and reduce clinical efficiency. Accelerated CMR reconstruction, which recovers high-quality images from undersampled k-space data, has become critical\cite{haldar2010compressed}\cite{zhu2018image}\cite{dar2020transfer}. Data-driven methods outperform traditional compressed sensing by enabling higher acceleration rates while preserving contrast-specific details and the spatiotemporal dynamics of cardiac imaging\cite{mardani2018deep}\cite{aggarwal2018modl}\cite{korkmaz2023self}.

Despite the success of data-driven approaches in CMR reconstruction, several key challenges remain, as outlined below.

\begin{itemize}
    \item Multi-task learning: Negative transfer\cite{Zhang_2023} and task interference\cite{gupta2024llmtaskinterferenceinitial} often degrade performance relative to single-task models.
    \item Out-of-distribution (OOD) generalization: Vendor/protocol/demographic shifts (artifacts, contrast) cause poor performance on unseen centers and rare pathologies \cite{wang2025one,nezhad2024generalizable,shorten2019survey}.
    \item Compute and memory: Large convolutional or deep unrolled models (DUM)\cite{sriram2020endtoendvariationalnetworksaccelerated} with unrolled data-consistency require high training time and GPU memory—especially for multi-coil and 3D/temporal data—hindering development and deployment.
\end{itemize}

We explored several methods that enhance a model’s representational capacity in multi-task settings without significantly increasing training cost. We also evaluated their feasibility on multi-domain datasets and on out-of-distribution (OOD) datasets.

\textbf{Our work includes the following components:}
\begin{itemize}
    \item We propose \textbf{Mixture-of-Experts for Deep Unrolled Model (Moero)}. Moero includes a minimal MoE implementation and specializes to heterogeneous distributions with minimal memory, multiplying parameters via parallel experts without increasing training-time GPU usage.
    \item We attempt to use a more robust dual-domain augmentation approach,  \textbf{DualSpace-CMR}, is proposed for MRI reconstruction. Physics-driven k-space augmentations simulate noise and undersampling\cite{desai2021vortex}, while image-domain augmentations preserve cardiac anatomy and motion\cite{fabian2021data}.
\end{itemize}

\section{Related Work}

\subsection{MRI image reconstruction}

In accelerated MRI, data are undersampled in k-space, leading to the inverse problem:

\begin{equation}
    \mathcal{F}_{\Omega} \, x = y_{\Omega},
\end{equation}

where \(x\) is the unknown image, \(\mathcal{F}_{\Omega}\) the partial Fourier transform, and \(y_{\Omega}\) the sampled measurements. To reduce aliasing artifacts, a regularized reconstruction is used:

\begin{equation}
    \hat{x} = \mathop{\mathrm{arg\,min}}\limits_{x \in \mathbf{C}^N} \frac{1}{2}||\mathcal{F}_{\Omega}x - y_{\Omega}||_{2}^{2} + \lambda R(x)
\end{equation}

where \(R(x)\) encodes prior knowledge, often implemented via a CNN \(\mathcal{D}_{\theta}\). 

Deep unrolled methods (DUM)\cite{sriram2020endtoendvariationalnetworksaccelerated} excel in MRI reconstruction, particularly for compressive sensing problems, due to their performance and interpretability \cite{NIPS2016_1679091c,xin2024rethinking}. These methods alternate between gradient descent and proximal mapping steps and are also effective in other inverse problems like image restoration and general reconstruction tasks.

Based on DUM\cite{sriram2020endtoendvariationalnetworksaccelerated}, Bingyu Xin et al.\cite{xin2024rethinking,xin2023fill} proposed PromptMR, which achieves an efficient end-to-end reconstruction method and introduces image-domain prompt into MRI reconstruction networks. The prompt is used to capture the global semantic information of the image.

\subsection{Vector Quantization}

VQ (Vector Quantization) \cite{oord2018neuraldiscreterepresentationlearning} maps continuous features to a finite set of representative codewords drawn from a learned codebook $\mathcal{C}=\{\mathbf{e}_k\}_{k=1}^{K}$ with $\mathbf{e}_k \in \mathbf{R}^d$. It reduces representational complexity by replacing each input vector $\mathbf{x}\in\mathbf{R}^d$ with its nearest codeword, while preserving salient structure.

\subsubsection{Nearest-neighbor Quantization}

Given $\mathcal{C}$ and $\mathbf{x}$:

\begin{equation}
    k^*=\mathop{\mathrm{arg\,min}}_{k\in\{1,\dots,K\}}\left\|\mathbf{x}-\mathbf{e}_k\right\|_2^2,\qquad
\hat{\mathbf{x}}=\mathbf{e}_{k^*}
\end{equation}

During training, VQ jointly learns the codebook and handles the non-differentiable assignment using the stop-gradient operator $\mathrm{sg}[\cdot]$ (or EMA-based updates). A typical objective combines a codebook (embedding) loss, a commitment loss, and an optional alignment term:

\begin{equation}
    \mathcal{L}_{\mathrm{VQ}} =
    \underbrace{\| \mathrm{sg}[z_e(x)] - e \|_2^2}_{\mbox{Codebook (embedding) loss}}
    + \beta \underbrace{\| z_e(x) - \mathrm{sg}[e] \|_2^2}_{\mbox{Commitment loss}}
    + \gamma \underbrace{\| z_e(x) - e \|_2^2}_{\mbox{Alignment}}.
\end{equation}

where $z_e(x)$ is the encoder output and $e=\mathbf{e}_{k^*}$.

\subsection{Mixture of Experts}

MoE (Mixture of Experts)\cite{6215056,zhou2022mixture} uses a gating network to route inputs to specialized experts and fuse their outputs, enabling diverse sub-distributions (e.g., undersampling, noise, motion, anatomy) to be handled by appropriate experts while scaling parameters with controlled compute and memory.

In recent years, MoE has been applied to MRI image super-resolution\cite{wang2025moediffsrmixtureexpertsguideddiffusion}, denoising\cite{Deng_2025}, and segmentation\cite{zhang2024foundationmodelbrainlesion}, and it shows strong potential as a foundation-modeRigid-body Motionl paradigm for MRI.

\section{Methods}

\subsection{DualSpace-CMR}

\subsubsection{Image-based Augmentations}

For reconstruction, image-domain augmentations must be propagated to k-space and coil sensitivities. We reconstruct the image from undersampled k-space using the sensitivity maps, apply the augmentation, then reapply the forward operator A to generate augmented k-space. Our image-level augmentations include:

\begin{itemize}
  \item Geometric transforms: flips, integer shifts, $90^\circ$/$180^\circ$/$270^\circ$ or arbitrary rotations, isotropic/anisotropic scaling.
  \item Elastic deformations.
\end{itemize}

\begin{figure}
\includegraphics[width=\textwidth]{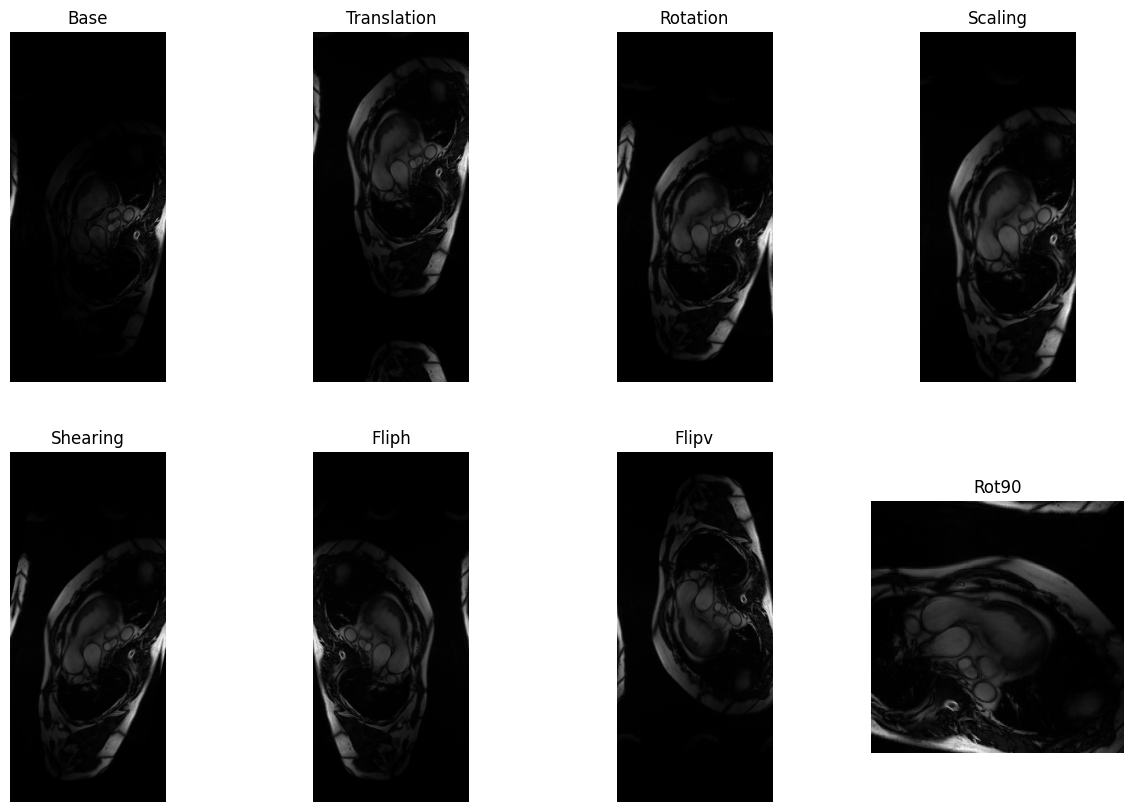}
\caption{Examples of image-based augmentations applied to a fully-sampled image.} \label{fig:image_augment}
\end{figure}

Convolutions are translation-equivariant but not rotation/scale invariant, so augmentation diversifies samples and reduces overfitting. Each geometric transform is applied jointly to image x and the coil-sensitivity maps to maintain forward-model consistency (Fig. \ref{fig:image_augment}).

\subsubsection{k-space Augmentations}

Instead of altering images, we inject controlled k-space artifacts to mimic realistic acquisition effects, focusing on thermal noise and rigid-body motion.

\paragraph{Thermal Noise}

Complex-valued thermal noise in MRI is well-approximated by additive white Gaussian noise on the k-space samples. Let y denote the fully-sampled k-space data. We generate a noise matrix $\mathcal{E}\sim\mathcal{CN}(0,\sigma^{2}\mathbf{I})$ and inject it into k-space:

\begin{equation}
    y_{\mathrm{noise}} = y + \mathcal{E}
\end{equation}

The standard deviation $\sigma$ is chosen from a discrete set $\sigma_{\mathrm{light}}$, $\sigma_{\mathrm{heavy}}$ that has been calibrated by a board-certified radiologist to reflect clinically observed SNR variations. After augmentation, the k-space is re-normalised so that each scan undergoes the same relative SNR change.

\paragraph{Rigid-body Motion}

Multi-shot sequences acquire different k-space lines in separate excitations. Rigid motion between shots introduces a 1-D translational displacement that manifests as a phase modulation along the phase-encoding direction.  
Let k=0,1,\dots,N-1 index the phase-encoding lines. For each shot we draw two independent uniform random variables $\phi_{\mathrm{odd}},\; \phi_{\mathrm{even}}\sim\mathcal{U}(-\pi,\pi)$ and apply the corresponding phase error: 

\begin{equation}
\psi(k)=
\left\{
\begin{array}{ll}
e^{\,j\phi_{\mathrm{odd}}}, & k\ \mbox{odd},\\[6pt]
e^{\,j\phi_{\mathrm{even}}}, & k\ \mbox{even}.
\end{array}
\right.
\end{equation}

The motion-corrupted k-space is then

\begin{equation}
    y_{\mathrm{motion}}(k,c)=\psi(k)\,y(k,c)\quad\forall c=1,\dots,C.
\end{equation}

\begin{figure}
\includegraphics[width=\textwidth]{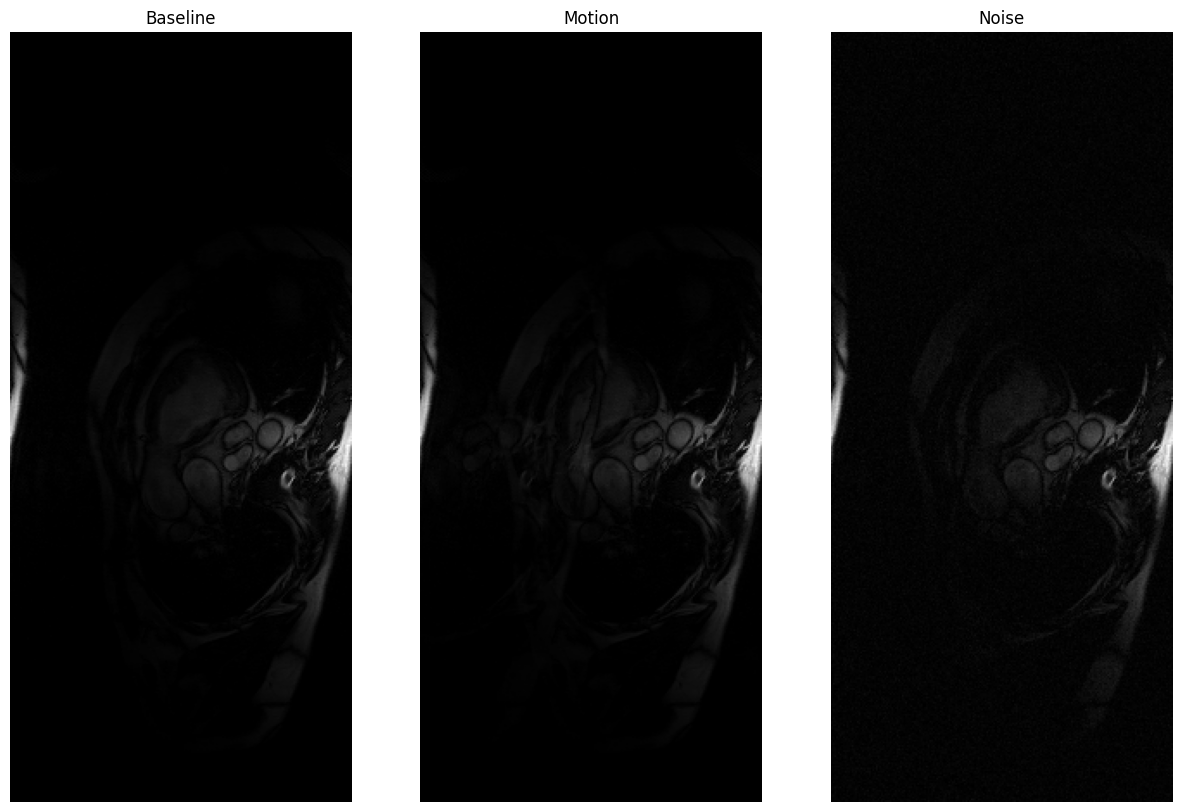}
\caption{Examples of k-space augmentations applied to a fully-sampled image.} \label{fig:kspace_augment}
\end{figure}

This closed-form model allows us to generate realistic ghosting artefacts without the need for navigator echoes or prolonged scan times.  We also provide an example of a k-space-augmented scan in Fig.\ref{fig:kspace_augment}.

\subsection{VQ-Prompt}

We introduce VQ-PromptBlock (Fig. \ref{fig:graph}d), a minimal VQ-VAE\cite{razavi2019generatingdiversehighfidelityimages} submodule designed to preserve high-quality features learned from the training data. Following PromptBlock\cite{xin2023fill}, we pass the vector-quantized codes through a lightweight multi-layer convolutional decoder and fuse them with the input via a residual connection. Since VQ-VAE\cite{razavi2019generatingdiversehighfidelityimages} can introduce blur in generative tasks, we place VQ-PromptBlock only at the U-Net bottleneck.

\subsection{Moero}

Inspired by MoE, we propose Moero, a multi-branch unrolled model where global representations guide branch routing, boosting expressivity with negligible training-time memory overhead. Moero comprises two components: Branch Grids and BranchNav.

\subsubsection{Branch Grids}

\begin{figure}
\includegraphics[width=\textwidth]{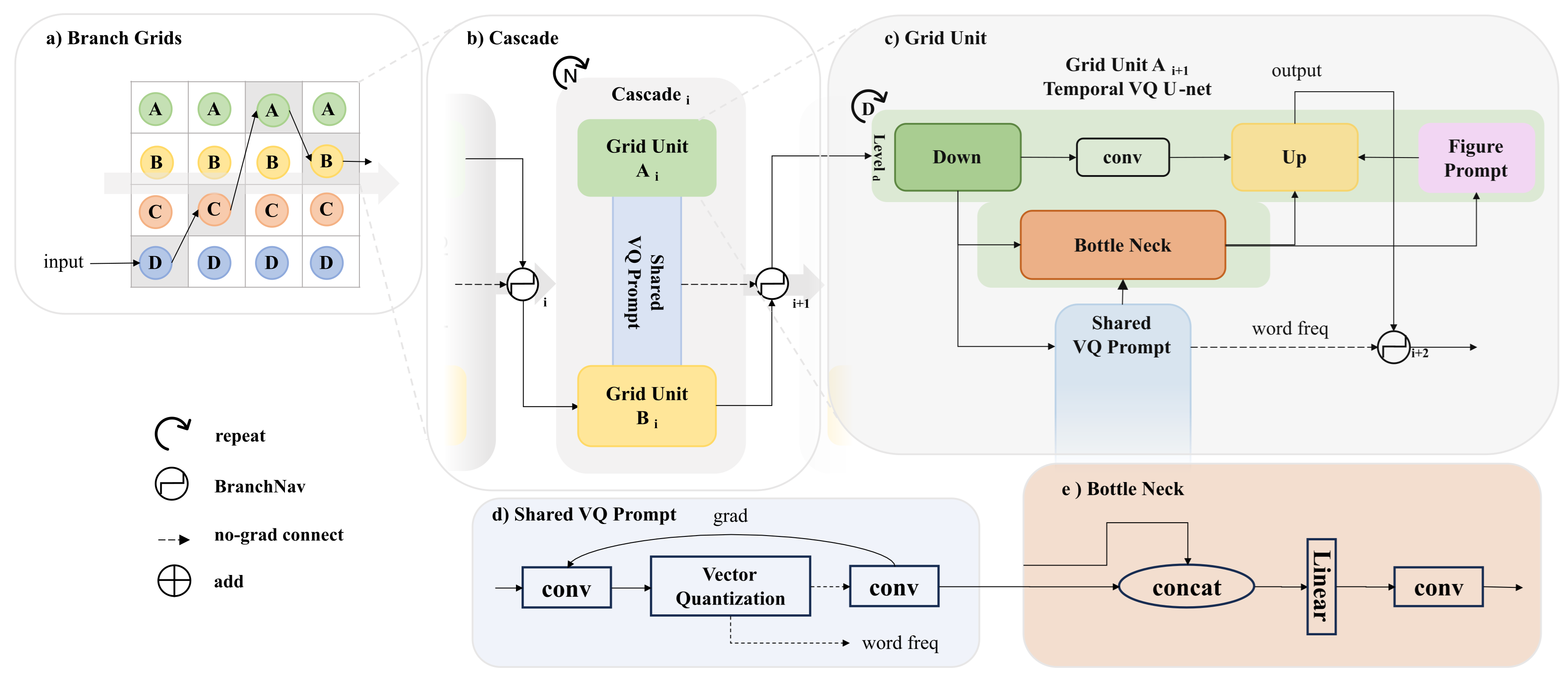}
\caption{Overview of Moero. We extend VarNet to a 2D Branch Grid (a) with axes \(N_{\mathrm{cascade}}\) and \(N_{\mathrm{branch}}\). Each cascade (b) contains a Grid Unit Network (c) and shares one embedding module (Shared VQ Prompt); the previous cascade’s embedding routes the next. The Grid Unit U-Net is augmented with Shared VQ Prompt (d), a bottleneck (e), and a Figure Prompt (following PromptMR \cite{xin2024rethinking,xin2023fill}). Shared VQ Prompt uses pixel-level features, whereas Figure Prompt operates in image space.} \label{fig:graph}
\end{figure}

Branch Grids adopt a Mixture-of-Experts (MoE) design: BranchNav performs sparse routing, activating a single path per forward pass, so parameters scale while compute/memory remain nearly constant. Each path specializes to a distribution mode, and each Grid Unit retains only the parameters needed for its assigned center, yielding compact, task-focused experts.

Each Grid Unit uses the Shared VQ Prompt embedding module to encode the image into a fixed-length token-frequency vector $\mathrm{word frequency} \in \mathbf{R}^{L_{\mathrm{word}}}$. $\mathrm{word frequency}$ counts the occurrences of each VQ codebook token (a normalized histogram) and serves as the expert-clustering feature. This vector routes the next cascade; for the first cascade, routing is determined by the $\mathrm{word frequency}$ produced by the sensitivity-map network.

The calculation of $\mathrm{word frequency}$ is as follows:

\begin{equation}
    \mathrm{word frequency}
= \frac{\mathrm{bincount}\!\left(\mathrm{VectorQuantize}(\mathbf{x})\mathrm{.index}\right)}{N_{\mathrm{pix}}}
\end{equation}

where $N_{pix}$ is the number of pixels in $x$.

\paragraph{Weights Sharing}

To ensure a consistent clustering feature space across branches, all experts within each cascade (each column in Fig. \ref{fig:graph}a) share the same VQ-PromptBlock and a common codebook. All other parameters are expert\-specific and not shared.

\subsubsection{BranchNav}

BranchNav serves as the sparse Mixture of Experts (MoE) router. It outputs a binary mask of length $N_{\mathrm{branch}}$, indicating which branch $i_{\mathrm{branch}}$ is activated. To initialize the Cluster Kernel, BranchNav buffers up to $N_{\mathrm{buffer}}$ word frequency vectors of length $L_{\mathrm{word}}$ and estimates the cluster centers. It also tracks the per‑center routing counts $C_{\mathrm{routing}}$. Once centers are initialized, BranchNav routes using the Cluster Kernel’s assignments; otherwise, it selects the least‑used branch via $\arg\min(C_{\mathrm{routing}})$. The computational rules are as follows:

\begin{equation}
\mbox{route}(\mbox{word frequency}) =
\left\{
\begin{array}{ll}
\mathop{\mathrm{arg\,min}}(C_{\mathrm{routing}}), &
\mbox{if } \frac{\max(C_{\mathrm{routing}})}{\min(C_{\mathrm{routing}})} > 3 \\[6pt]
& \mbox{or kernel is not initialized}, \\[6pt]
\mathrm{Kernel}(\mbox{word frequency}), & \mbox{Otherwise}.
\end{array}
\right.
\end{equation}

BranchNav prioritizes balanced routing: if an expert has abnormally few routed samples, it is selected first; otherwise, BranchNav follows the clustering assignment. To ensure that the expert assignments for the same sample remain stable, BranchNav does not buffer any samples during the first epoch.

To simplify training, we adopt a non-learned clustering core using \textbf{k-means}. To mitigate sparsity in high-dimensional spaces, we employ \textbf{Cosine Similarity} as the distance metric.

\section{Experiments}

\subsection{Data}

We use CMRxRecon25: 1,384 k-space volumes spanning cine, T1/T2 mapping, LGE, phase-contrast, T1w/T2w, perfusion, and dark-blood. Data were acquired on Siemens/GE/UIH scanners at six sites with harmonized protocols, covering 2/3/4-chamber LAX, SAX stacks, LVOT, and aortic views. Training uses fully sampled data; inference uses undersampled data from two additional centers with pathologies to assess generalization to unseen hardware and anomalie.

\subsection{Training Details}

We resample the dataset using an adaptive sample-balancing sampler. The draw count for each sample is determined by the following weight:

\begin{equation}
    \mathrm{weight}\leftarrow \mathrm{weight}\cdot\Bigg[\,1+\Big(\frac{N/n_r}{c_r\!\big(f_r(i)\big)}-1\Big)\cdot 0.8\,\Bigg].
\end{equation}

where $N$ is the total number of samples, $n_r$ is the number of groups under category $r$, $f_r(i)$ denotes the group of sample $i$, and $c_r(g)$ denotes the size (count) of group $g$.

During each training iteration, one undersampling mask \{Uniform, kt-Gaussian, kt-Radial\} is randomly generated with acceleration \{4, 8, 10, 12, 16, 20, 24\}. All masks use 20 autocalibration (ACS) lines.

We conduct ablation experiments. The baseline and augmentation settings strictly follow the original PromptMR-plus \cite{xin2024rethinking}. We evaluate two variants, baseline with VQPromptBlock, and Moero, against the baseline. To control variables, all models use a 3 levels U-Net with channels [72, 96, 120] and 12 cascades, matching the baseline. The VQ-Prompt codebook size is set to the smallest value that avoids large VQ-loss oscillations during training; here, it is $1024$.

All experiments were run on 8 NVIDIA-A100 GPUs. Data-augmentation experiments were trained for 12 epochs. Model-augmentation experiments were trained for 12, 18, and 24 epochs to assess performance at different stages of overfitting. Due to limited resources, we stopped extending experiments once pronounced overfitting was observed.

\subsection{Results and Discussions}

\begin{table}
\caption{Results of data enhancement experiments}
\label{tab:restab1}
\begin{tabular*}{\linewidth}{@{\extracolsep{\fill}}|l|l|l|l|}
\hline
Method & SSIM & PSNR & NMSE \\
\hline
PromptMR+ (baseline) & 0.836 & 30.413 & 0.028 \\
PromptMR+ with kspace aug & 0.824 & 29.803 & 0.030 \\
PromptMR+ with image aug & 0.816 & 29.795 & 0.043 \\
\hline
\end{tabular*}
\end{table}

\begin{table}
\caption{Results of model-variant experiments (SSIM, memory, and iteration rate). rare denotes classes with few samples, and fresh denotes sample categories that were not used during training.}
\label{tab:restab2}
\begin{tabular*}{\linewidth}{@{\extracolsep{\fill}}|l|l|l|l|l|l|}
\hline
Method & SSIM-total & SSIM-rare & SSIM-fresh & \begin{tabular}[t]{@{}l@{}}Memory\\avg+peak\\(train)\end{tabular} & Iteration/s \\
\hline
\begin{tabular}[t]{@{}l@{}}PromptMR+\\(baseline, epoch12)\end{tabular} & 0.836 & 0.739 & 0.786 & 33.67+1.95 (GB) & 1.27 \\
\begin{tabular}[t]{@{}l@{}}baseline + VQPrompt\\(epoch12)\end{tabular} & 0.8442 & 0.7565 & 0.7989 & 33.95+2.42 (GB) & 1.18 \\
\begin{tabular}[t]{@{}l@{}}baseline + VQPrompt\\(epoch18)\end{tabular} & 0.8445 & 0.7595 & 0.801 & $\sim$ & $\sim$ \\
\begin{tabular}[t]{@{}l@{}}Moero\\(branch2, epoch12)\end{tabular} & 0.828 & 0.733 & 0.78 & 34.18+2.57 (GB) & 0.98 \\
\begin{tabular}[t]{@{}l@{}}Moero\\(branch4, epoch12)\end{tabular} & 0.806 & 0.71 & 0.753 & 36.62+3.26 (GB) & 0.86 \\
\begin{tabular}[t]{@{}l@{}}Moero\\(branch4, epoch18)\end{tabular} & 0.826 & 0.737 & 0.779 & $\sim$ & $\sim$ \\
\begin{tabular}[t]{@{}l@{}}Moero\\(branch4, epoch24)\end{tabular} & 0.834 & 0.744 & 0.786 & $\sim$ & $\sim$ \\
\hline
\end{tabular*}
\end{table}

In small dataset testing, the combined augmentation of motion and noise on k-space datasets achieves better results than single noise augmentation, single motion augmentation, or image augmentation. However, on the large-scale benchmark (Tab. \ref{tab:restab1}) the same strategy degrades the metrics. Additionally, we observed that the data-augmentation ratio strongly influences test performance: if set too low, gains are negligible; if set too high, severe distortion degrades training quality. Careful tuning of this parameter is therefore critical.

We plan to conduct some research on data generation in the future, generating data with cross-center characteristics, and then using this data in subsequent deep learning model training. The trained model is theoretically expected to have a certain cross-center inference capability.

In the model-augmentation experiments (Tab. \ref{tab:restab2}), both the baseline and the VQPrompt-augmented variant plateau after epoch 12, with virtually no further reduction in loss. In contrast, Moero continues to exhibit a steady, albeit slow, decline. Given its prohibitive runtime (over 4 days) and limited experimental resources, we stopped the experiment at epoch 24.

Additionally, in our experiments (Tab. \ref{tab:restab2}), the overfitting caused by MoE did not materially degrade generalization on few-shot or out-of-distribution samples, and performance remained in line with the baseline trend rather than deviating from it.

Introducing the VQPromptBlock yields a slight performance improvement, suggesting that vector quantization is promising for MRI reconstruction. However, the current evidence is limited; additional ablation studies and cross-validation are needed to rule out noise and confounding effects.

Because Moero uses a sparse MoE strategy and training-time memory in image generation is dominated by the autograd graph, its GPU memory usage increases only modestly relative to the baseline. However, training sparse networks in \textbf{PyTorch} requires detecting unused parameters on every iteration, and the resulting gradient synchronization overhead reduces iteration throughput, even for experts that are not activated by the current batch. Additionally, with 2 experts, Moero requires roughly 1.5x as many iterations as the baseline; with 4 experts, about 2x. While it is encouraging that iteration count does not scale linearly with the number of experts, the wall-clock time remains difficult to justify under our current constraints. Therefore, improving the efficiency of MoE training is a promising direction for Moero.

\section{Conclusion}

We explored two complementary directions for multi-contrast CMR reconstruction: data-space augmentation (DualSpace-CMR) and model-capacity scaling (VQ-Prompt, Moero). On small datasets, k-space motion+noise helped; on the large benchmark it hurt, underscoring sensitivity to augmentation ratio and schedule. VQ-Prompt delivered small, consistent gains with negligible memory; Moero improved steadily after epoch 12 and retained baseline-like few-shot/OOD generalization despite mild overfitting. Memory overhead was modest, but sparse routing slowed PyTorch throughput, making wall-clock time the bottleneck.

%
%
%
\bibliographystyle{splncs04}
\bibliography{refs}

@article{wang2025one,
  title={One for multiple: Physics-informed synthetic data boosts generalizable deep learning for fast MRI reconstruction},
  author={Wang, Zi and Yu, Xiaotong and Wang, Chengyan and Chen, Weibo and Wang, Jiazheng and Chu, Ying-Hua and Sun, Hongwei and Li, Rushuai and Li, Peiyong and Yang, Fan and others},
  journal={Medical Image Analysis},
  pages={103616},
  year={2025},
  publisher={Elsevier}
}

@inproceedings{nezhad2024generalizable,
  title={Generalizable deep mri reconstruction with cross-site data synthesis},
  author={Nezhad, Valiyeh Ansarian and Elmas, G{\"o}kberk and Arslan, Fuat and Kaba{\c{s}}, Bilal and {\c{C}}ukur, Tolga},
  booktitle={2024 32nd Signal Processing and Communications Applications Conference (SIU)},
  pages={1--4},
  year={2024},
  organization={IEEE}
}

@inproceedings{fabian2021data,
  title={Data augmentation for deep learning based accelerated MRI reconstruction with limited data},
  author={Fabian, Zalan and Heckel, Reinhard and Soltanolkotabi, Mahdi},
  booktitle={International Conference on Machine Learning},
  pages={3057--3067},
  year={2021},
  organization={PMLR}
}

@article{desai2021vortex,
  title={Vortex: Physics-driven data augmentations using consistency training for robust accelerated mri reconstruction},
  author={Desai, Arjun D and Gunel, Beliz and Ozturkler, Batu M and Beg, Harris and Vasanawala, Shreyas and Hargreaves, Brian A and R{\'e}, Christopher and Pauly, John M and Chaudhari, Akshay S},
  journal={arXiv preprint arXiv:2111.02549},
  year={2021}
}

@article{haldar2010compressed,
  title={Compressed-sensing MRI with random encoding},
  author={Haldar, Justin P and Hernando, Diego and Liang, Zhi-Pei},
  journal={IEEE transactions on Medical Imaging},
  volume={30},
  number={4},
  pages={893--903},
  year={2010},
  publisher={IEEE}
}

@article{zhu2018image,
  title={Image reconstruction by domain-transform manifold learning},
  author={Zhu, Bo and Liu, Jeremiah Z and Cauley, Stephen F and Rosen, Bruce R and Rosen, Matthew S},
  journal={Nature},
  volume={555},
  number={7697},
  pages={487--492},
  year={2018},
  publisher={Nature Publishing Group UK London}
}

@article{dar2020transfer,
  title={A transfer-learning approach for accelerated MRI using deep neural networks},
  author={Dar, Salman Ul Hassan and {\"O}zbey, Muzaffer and {\c{C}}atl{\i}, Ahmet Burak and {\c{C}}ukur, Tolga},
  journal={Magnetic resonance in medicine},
  volume={84},
  number={2},
  pages={663--685},
  year={2020},
  publisher={Wiley Online Library}
}

@article{mardani2018deep,
  title={Deep generative adversarial neural networks for compressive sensing MRI},
  author={Mardani, Morteza and Gong, Enhao and Cheng, Joseph Y and Vasanawala, Shreyas S and Zaharchuk, Greg and Xing, Lei and Pauly, John M},
  journal={IEEE transactions on medical imaging},
  volume={38},
  number={1},
  pages={167--179},
  year={2018},
  publisher={IEEE}
}

@article{aggarwal2018modl,
  title={MoDL: Model-based deep learning architecture for inverse problems},
  author={Aggarwal, Hemant K and Mani, Merry P and Jacob, Mathews},
  journal={IEEE transactions on medical imaging},
  volume={38},
  number={2},
  pages={394--405},
  year={2018},
  publisher={IEEE}
}

@inproceedings{korkmaz2023self,
  title={Self-supervised MRI reconstruction with unrolled diffusion models},
  author={Korkmaz, Yilmaz and Cukur, Tolga and Patel, Vishal M},
  booktitle={International Conference on Medical Image Computing and Computer-Assisted Intervention},
  pages={491--501},
  year={2023},
  organization={Springer}
}

@article{shorten2019survey,
  title={A survey on image data augmentation for deep learning},
  author={Shorten, Connor and Khoshgoftaar, Taghi M},
  journal={Journal of big data},
  volume={6},
  number={1},
  pages={1--48},
  year={2019},
  publisher={Springer}
}

@inproceedings{xin2024rethinking,
  title={Rethinking deep unrolled model for accelerated MRI reconstruction},
  author={Xin, Bingyu and Ye, Meng and Axel, Leon and Metaxas, Dimitris N},
  booktitle={European Conference on Computer Vision},
  pages={164--181},
  year={2024},
  organization={Springer}
}

@inproceedings{NIPS2016_1679091c,
 author = {yang, yan and Sun, Jian and Li, Huibin and Xu, Zongben},
 booktitle = {Advances in Neural Information Processing Systems},
 editor = {D. Lee and M. Sugiyama and U. Luxburg and I. Guyon and R. Garnett},
 pages = {},
 publisher = {Curran Associates, Inc.},
 title = {Deep ADMM-Net for Compressive Sensing MRI},
 volume = {29},
 year = {2016}
}

@misc{oord2018neuraldiscreterepresentationlearning,
      title={Neural Discrete Representation Learning}, 
      author={Aaron van den Oord and Oriol Vinyals and Koray Kavukcuoglu},
      year={2018},
      eprint={1711.00937},
      archivePrefix={arXiv},
      primaryClass={cs.LG},
      url={https://arxiv.org/abs/1711.00937}, 
}

@article{xin2023fill,
  title={Fill the K-Space and Refine the Image: Prompting for Dynamic and Multi-Contrast MRI Reconstruction},
  author={Xin, Bingyu and Ye, Meng and Axel, Leon and Metaxas, Dimitris N},
  journal={arXiv preprint arXiv:2309.13839},
  year={2023}
}

@article{Zhang_2023,
   title={A Survey on Negative Transfer},
   volume={10},
   ISSN={2329-9274},
   url={http://dx.doi.org/10.1109/JAS.2022.106004},
   DOI={10.1109/jas.2022.106004},
   number={2},
   journal={IEEE/CAA Journal of Automatica Sinica},
   publisher={Institute of Electrical and Electronics Engineers (IEEE)},
   author={Zhang, Wen and Deng, Lingfei and Zhang, Lei and Wu, Dongrui},
   year={2023},
   month=feb, pages={305–329} }

@misc{gupta2024llmtaskinterferenceinitial,
      title={LLM Task Interference: An Initial Study on the Impact of Task-Switch in Conversational History}, 
      author={Akash Gupta and Ivaxi Sheth and Vyas Raina and Mark Gales and Mario Fritz},
      year={2024},
      eprint={2402.18216},
      archivePrefix={arXiv},
      primaryClass={cs.CL},
      url={https://arxiv.org/abs/2402.18216}, 
}

@misc{sriram2020endtoendvariationalnetworksaccelerated,
      title={End-to-End Variational Networks for Accelerated MRI Reconstruction}, 
      author={Anuroop Sriram and Jure Zbontar and Tullie Murrell and Aaron Defazio and C. Lawrence Zitnick and Nafissa Yakubova and Florian Knoll and Patricia Johnson},
      year={2020},
      eprint={2004.06688},
      archivePrefix={arXiv},
      primaryClass={eess.IV},
      url={https://arxiv.org/abs/2004.06688}, 
}

@article{zhou2022mixture,
  title={Mixture-of-experts with expert choice routing},
  author={Zhou, Yanqi and Lei, Tao and Liu, Hanxiao and Du, Nan and Huang, Yanping and Zhao, Vincent and Dai, Andrew M and Le, Quoc V and Laudon, James and others},
  journal={Advances in Neural Information Processing Systems},
  volume={35},
  pages={7103--7114},
  year={2022}
}

@ARTICLE{6215056,
  author={Yuksel, Seniha Esen and Wilson, Joseph N. and Gader, Paul D.},
  journal={IEEE Transactions on Neural Networks and Learning Systems}, 
  title={Twenty Years of Mixture of Experts}, 
  year={2012},
  volume={23},
  number={8},
  pages={1177-1193},
  doi={10.1109/TNNLS.2012.2200299}}

@misc{wang2025moediffsrmixtureexpertsguideddiffusion,
      title={MoEDiff-SR: Mixture of Experts-Guided Diffusion Model for Region-Adaptive MRI Super-Resolution}, 
      author={Zhe Wang and Yuhua Ru and Aladine Chetouani and Fang Chen and Fabian Bauer and Liping Zhang and Didier Hans and Rachid Jennane and Mohamed Jarraya and Yung Hsin Chen},
      year={2025},
      eprint={2504.07308},
      archivePrefix={arXiv},
      primaryClass={eess.IV},
      url={https://arxiv.org/abs/2504.07308}, 
}

@inproceedings{Deng_2025,
   title={Sparse Mixture-of-Experts for Non-Uniform Noise Reduction in MRI Images},
   url={http://dx.doi.org/10.1109/WACVW65960.2025.00036},
   DOI={10.1109/wacvw65960.2025.00036},
   booktitle={2025 IEEE/CVF Winter Conference on Applications of Computer Vision Workshops (WACVW)},
   publisher={IEEE},
   author={Deng, Zeyun and Campbell, Joseph},
   year={2025},
   month=feb, pages={260–268} }

@misc{zhang2024foundationmodelbrainlesion,
      title={A Foundation Model for Brain Lesion Segmentation with Mixture of Modality Experts}, 
      author={Xinru Zhang and Ni Ou and Berke Doga Basaran and Marco Visentin and Mengyun Qiao and Renyang Gu and Cheng Ouyang and Yaou Liu and Paul M. Matthew and Chuyang Ye and Wenjia Bai},
      year={2024},
      eprint={2405.10246},
      archivePrefix={arXiv},
      primaryClass={eess.IV},
      url={https://arxiv.org/abs/2405.10246}, 
}

@misc{razavi2019generatingdiversehighfidelityimages,
      title={Generating Diverse High-Fidelity Images with VQ-VAE-2}, 
      author={Ali Razavi and Aaron van den Oord and Oriol Vinyals},
      year={2019},
      eprint={1906.00446},
      archivePrefix={arXiv},
      primaryClass={cs.LG},
      url={https://arxiv.org/abs/1906.00446}, 
}
\end{document}